\documentclass[useAMS,eqsecnum,usenatbib,onecolumn]{mn2e}

\usepackage{graphicx}
\usepackage{dcolumn}
\usepackage{bm}
\usepackage{natbib}
\bibpunct{(}{)}{,}{a}{}{,}
\begin{document}


\title{Universality in Quasi-normal Modes of Neutron Stars}

\author[L.K.~Tsui and P.T.~Leung]{L.K.~Tsui and P.T.~Leung\thanks{Email:
ptleung@phy.cuhk.edu.hk}\\
Physics Department, The Chinese University of Hong Kong, Shatin,
Hong Kong SAR, China.}

\date{\today}
\pagerange{\pageref{firstpage}--\pageref{lastpage}} \pubyear{2004}

\label{firstpage}
\def\tomega{ \tilde{\omega} }
\def\tOmega{ \tilde{\Omega} }
\def\tr{ \tilde{r} }
\def\tx{ \tilde{r}_* }
\def\tV{ \tilde{V} }
\def\tpsi{ \tilde{\psi} }
\def\trho{ \tilde{\rho} }
\def\tP{ \tilde{P} }
\def\tR{ \tilde{R} }
\def\tX{ \tilde{R}_* }
\def\tm{ \tilde{m} }
\def\tphi{ \tilde{\nu} }
\def\tlam{ \tilde{\lambda}}
\def\tepsilon { \tilde{\epsilon}}
\def\tHO{ \tilde{H}_0}
\def\tHI{ \tilde{H}_1}
\def\tK{ \tilde{K}}
\def\tW{ \tilde{W}}
\def\tX{ \tilde{X}}
\def\tV{ \tilde{V}}
\def\tgamma{ \tilde{\gamma}}
\def\ta{ \tilde{a}}
\def\tb{ \tilde{b}}
\def\tg{ \tilde{g}}
\def\th{ \tilde{h}}
\def\tk{ \tilde{k}}
\def\a{ {\rm a}}
\def\c{ {\rm c}}
\def\e{ {\rm e}}
\def\p{ {\rm p}}
\def\f{ {\rm f}}
\def\d{ {\rm d}}
\def\i{ {\rm i}}
\def\r{ {\rm r}}
\maketitle
\begin{abstract}
We study the universality in gravitational waves emitted from
non-rotating neutron stars characterized by different equations of
state (EOS). We find that the quasi-normal mode frequencies of
such waves, including the $w$-modes and the $f$-mode, display
similar universal scaling behaviours that hold for most EOS. Such
behaviours are shown to stem from the mathematical structure of
the axial and the polar gravitational wave equations, and the fact
that the mass distribution function can be approximated by a
cubic-quintic polynomial in radius. As a benchmark for other
realistic neutron stars, a simple model of neutron stars is
adopted here to reproduce the pulsation frequencies and the
generic scaling behaviours mentioned above with good accuracy.
\end{abstract}

\section{Introduction}
The search for gravitational wave has been the goal of endeavour
for generations of physicists since its existence was predicted in
the theory of general relativity \citep[see, e.g. a review by
Thorne in][chap.~9]{Thorne_rev}. In the beginning of a new
millennium, the breakthroughs in technology and endless efforts of
researchers strongly convince us that the first detection of
gravitational waves emitted from violent stellar collapses will be
realized within several years \citep[see e.g.][and references
therein]{Hughes_03,Grishchuk}. Gravitational wave detectors of
different designs, including ground-based interferometers (e.g.
LIGO and VIRGO),  and resonant antennas (e.g. EXPLOPER and NIBOE),
are either already in operation or in their final testing stages
\citep{Hughes_03,Grishchuk}. These detectors, together with the
space-based interferometers LISA, which is currently under
construction and is expected to be launched around 2011, can
detect gravitational waves in different frequency ranges. As a
consequence, gravitational-wave astronomy that enables researchers
to probe astronomical activities at extremely large distances from
gravitational waves detected emerges as a new subject of interest
\citep{Hughes_03,Grishchuk}.

Among all the possible sources of gravitational waves, pulsating
neutron stars may probably be the most interesting one
\citep{Thorne67}. It is believed that gravitational waves emitted
in binary neutron mergers are plausibly detectable with the
current technology and the frequency of detection could be as high
as several hundreds per year  after upgrading the present
detectors \citep{Hughes_03,Belczynski:2001uc}. Besides, waves
generated in the formation of neutron stars via stellar core
collapses might also be strong enough for detection provided that
the collapse is sufficiently asymmetric
\citep{Fryer:2001zw,Lindblom:1998wf}. Most interestingly, such
gravitational waves are likely to carry the information about the
internal structure of neutron stars from which they are emitted.
As the structure of a neutron star, such as its radius and density
distribution, certainly reflects the properties of nuclear and
quark matters, a close examination of the wave signals is deemed
rewarding
\citep{Andersson_1996,Andersson1998,Ferrari,Kokkotas_2001}.

Gravitational waves are commonly analyzed in terms of quasi-normal
modes (QNMs), which properly describe damped harmonic pulsations.
A QNM oscillation has a time dependence $\exp(i\omega t)$ and is
characterized by a complex eigenfrequency
$\omega=\omega_\r+\i\omega_\i$
\citep{Press_1971,Leaver_1986,Ching,Kokkotas_rev}. The QNM
frequencies of gravitational waves of a neutron star with a fixed
mass $M$ are generally model-sensitive and significantly depend on
the EOS adopted in the stellar model. However, some universal
behaviours in the frequency $\omega_\r$ and the damping time
$\tau\equiv 1/\omega_\i$ of the leading gravitational wave
$w$-mode of non-rotating neutron stars have recently been observed
\citep{Andersson1998,Ferrari}, which can be summarized by the
following pair of formulas:
\begin{eqnarray}
\omega_\r & \approx & \frac{1}{R}\left[a_\r
\left(\frac{M}{R}\right)+b_\r
\right] ,\label{BBF1}\\
\frac{1}{\tau}& \approx & \frac{1}{M}\left[a_\i
\left(\frac{M}{R}\right)^2+b_\i\left(\frac{M}{R}\right)+c_\i
\right].\label{BBF2}
\end{eqnarray}
Here $R$ is the radius of the star, $a_\r$, $b_\r$, $a_\i$, $b_\i$
and $c_\i$ are model-independent constants determined from curve
fitting. It is worthy of remark that such simple scaling formulas
apply to both axial and polar gravitational wave $w$-modes of
neutron stars \citep{Andersson1998,Ferrari}.

On the other hand, QNM frequencies of the fluid $f$-mode
 are discovered to follow approximately another pair of universal
formulas \citep{Andersson1998}:
\begin{eqnarray}
\omega_\r & \approx & \alpha_\r
\left(\frac{M}{R^3}\right)^{1/2}+\beta_\r
,\label{BBF_f1}\\
\frac{1}{\tau}& \approx &
\frac{M^3}{R^4}\left[\alpha_\i\left(\frac{M}{R}\right)+\beta_\i
\right],\label{BBF_f2}
\end{eqnarray}
where again $\alpha_\r$, $\beta_\r$, $\alpha_\i$, and $\beta_\i$
 are model-independent constants chosen to yield best fit. Based on
 the above-mentioned universal behaviours, it has been shown that
the radius and the mass of a neutron star can be inferred from the
pulsation frequencies of its fundamental fluid $f$-mode and the
first $w$-mode, and the equation of state (EOS) can in turn be
identified \citep{Andersson1998,Ferrari}.

These universal behaviours are intriguing in their own right. Why
can Eqs.~(\ref{BBF1})-(\ref{BBF_f2}) successfully reproduce the
QNM frequencies for stars with different EOS? What are their
physical interpretations? Is there any exception to them? In this
paper, we seek the physical mechanism underlying the
above-mentioned universality in the gravitational wave of
non-rotating neutron stars. By properly scaling the axial and
polar wave equations, we show that the scaled complex
eigenfrequencies $M\omega$ of axial and polar $w$-modes, and
$f$-mode oscillations, to a good approximation, depend only on the
compactness $M/R$. Moreover, to interpret such EOS-independent
generic behaviour, we use a simple stellar model --- the Tolman
VII model (TVIIM) ---   to approximate the mass distribution
inside realistic neutron stars \citep{Tolman:1939jz}, and find
that TVIIM is indeed a good approximation. We then look for the
QNMs of TVIIM with the scaled wave equation. Interestingly enough,
such QNMs manifestly depend only on the compactness of the star
and, in addition, demonstrate the scaling behaviour discovered in
realistic neutron stars. Therefore, the universality in the QNMs
($w$-mode and $f$-mode) is ascribable to the fact that that the
mass distribution of most realistic neutron stars can be well
approximated by TVIIM.

The organization of our paper is as follows. In Sect.~2, we show
how the frequencies (including the real and the imaginary parts)
of QNMs of neutron stars with different EOS can be captured by a
single formula. In Sect.~3 we briefly review the equilibrium
configuration of a neutron star and introduce TVIIM as an
approximation of realistic neutron stars. Sections 4 and 5 study
the scaling behaviours of the axial and the polar pulsation modes,
respectively. In Sect.~6 we discuss potential application of our
finding and consider some possible exceptional cases where obvious
deviations from the universality are expected. We then conclude
our paper in Sect.~7. Unless otherwise stated, geometrized units
in which $G=c=1$ are adopted in the following discussion.

\section{Generic behaviour of QNMs}
The theory of non-radial oscillations of relativistic stars was
pioneered by \citet{Thorne67}. Later, \citet{Lindblom_1983}, and
\citet{Chandrasekhar1} further simplified the equations describing
gravitational wave generation and fluid motion, which will be
detailed in the later part of this paper. To display the
universality in the oscillation modes, we numerically integrate
the gravitational wave equations for neutron stars, and impose the
outgoing wave boundary condition at spatial infinity to locate
relevant QNMs. Eight different EOS, including models A
\citep{modelA} and C \citep{modelC} proposed by Pandharipande,
three models (AU, UU and UT) proposed by \citet{AU}, models APR1
and APR2 proposed by \citet{APR}, and model GM${24}$
\citep[p.~244]{ComStar} are considered in our papers. Figures
\ref{ldw_a}, \ref{ldw_p} and \ref{f_mode} show the relation
between $M \omega$ and the compactness $M/R$ for the first axial
$w$-mode, the first polar $w$-mode and the $f$-mode, respectively.
It is clearly shown that both $M \omega_\r$ and $M \omega_\i$
depend on the compactness in a universal way. In particular, as
shown by the solid line in these figures, a quadratic function in
$M/R$, namely:
\begin{eqnarray}\label{regression}
M \omega = a
\left(\frac{M}{R}\right)^2+b\left(\frac{M}{R}\right)+c \label{TL}
\end{eqnarray}
successfully reproduces the dependence on the compactness in each
case. Here $a$, $b$ and $c$ are complex constants determined by
regression and their values for individual cases are tabulated in
Table 1 for reference. While reproducing the observed universality
of the $w$-mode as described in (\ref{BBF1}) and (\ref{BBF2}),
Eq.~(\ref{TL}) is different from (\ref{BBF_f1}) and (\ref{BBF_f2})
discovered previously \citep{Andersson1998}. In fact, for nonzero
$\beta_\r$, Eq.~(\ref{BBF_f1}) leads to explicit mass-dependence
of $M\omega_\r$, which is not observed in our study. On the other
hand, we note that $p$-mode QNM frequencies (not shown in the
present paper), especially the imaginary part, do not demonstrate
analogous behaviour. However, as shown in Fig.~\ref{ldw_a2} and
\ref{ldw_p2}, such universality also exists in non-leading axial
and polar $w$-modes.
\begin{table}
\caption{Values of  $a$, $b$ and $c$ in (\ref{TL}).}
\begin{tabular}{llr@{}c@{}lr@{}c@{}lr@{}c@{}l}
\hline Wave type & Data &&{\emph{a}}&&&{\emph{b}}&&&{\emph{c}}&
\\\hline
1st axial $w$-mode & realistic stars & $-$3.9&$-$&i5.6& 2.8&+&i1.6 & $-$0.030&+&i0.12 \\
                   & TVIIM             & $-$4.4&$-$&i6.3& 3.1&+&i1.9 & $-$0.072&+&i0.098 \\
2nd axial $w$-mode & realistic stars & $-$13&$-$&i8.1& 6.1&+&i2.8 & 0.034&+&i0.11 \\
                   & TVIIM             & $-$14&$-$&i9.3& 6.7&+&i3.3 & $-$0.049&+&i0.060 \\
1st polar $w$-mode & realistic stars & $-$8.2&$-$&i10& 3.9&+&i3.3 & 0.055&+&i0.043 \\
                   & TVIIM             & $-$8.6&$-$&i11& 4.2&+&i3.6 & 0.0080&+&i0.014 \\
2nd polar $w$-mode & realistic stars & $-$18&$-$&i8.6& 8.0&+&i3.1 & 0.031&+&i0.11 \\
                   & TVIIM             & $-$20&$-$&i9.9& 8.8&+&i3.6 & $-$0.077&+&i0.058 \\
polar $f$-mode & realistic stars     & 0.15&$-$&i5.8\,E-4& 0.56&+&i6.7\,E-4 & $-$0.020&$-$&i6.2\,E-5 \\
                   & TVIIM             & 0.14&$-$&i6.9\,E-4& 0.60&+&i7.2\,E-4 & $-$0.027&$-$&i6.8\,E-5 \\
\hline
\end{tabular}
\end{table}

The neutron stars considered here are constructed from nuclear
matters with different compositions (e.g. pure neutron matter and
hyperon matter) and different stiffness. It is well known that the
mass-radius curve for neutron stars reveals marked EOS dependence.
Yet the  scaled QNM eigenfrequency, $M \omega$, can display an
EOS-independent generic behaviour. Two questions arise naturally
from our finding. First, why is the scaled eigenfrequency, $M
\omega$, EOS-insensitive? Second, does the universal curve in
Eq.~(\ref{TL}) has any physical meaning? In the ensuing discussion
we will work out the answers to these questions.
\section{Equilibrium configuration and TVIIM}
To look for a simple yet robust model for neutron stars, we first
briefly review the equilibrium configuration of non-rotating
neutron stars. In spherical coordinates, $(t,r,\theta,\varphi)$,
the geometry of spacetime around a non-rotating neutron star is
given by the line element:
\begin{equation}
\d s^2=-\e^{\nu(r)}\d t^2+\e^{\lambda(r)}\d
r^2+r^2(\d\theta^2+\sin^2\theta \d\varphi^2).
\end{equation}
where the metric coefficient $\e^{\lambda(r)}$ is determined by
the mass distribution function $m(r)$, the mass inside
circumferential radius $r$, as follows:
\begin{eqnarray}
    \e^{-\lambda(r)}&=&1-\frac{2m(r)}{r}. \label{g1}
\end{eqnarray}
On the other hand, the metric coefficient $\e^{\nu(r)}$, the
density of mass-energy, $\rho(r)$, and the pressure $P(r)$ obey
the Tolman-Oppenheimer-Volkoff (TOV) equations
\citep{Tolman:1939jz,Oppenheimer:1939ne}:
\begin{eqnarray}
\frac{\d\nu}{\d r} &=& \frac{2m+8\pi r^3 P}{r(r-2m)} ,\label{g2}\\
\frac{\d m}{\d r} &=& 4\pi r^2\rho, \label{TOV1}\\
\frac{\d P}{\d r} &=& -\frac{1}{2}(\rho+P)\frac{\d\nu}{\d
 r}.\label{TOV2}
\end{eqnarray}
These four equations, together with an EOS, $P=P(\rho)$, and the
following boundary conditions at the surface of the star where
$r=R$:
\begin{equation}\label{P_boundary}
    P(r=R)=0,
\end{equation}
\begin{equation}\label{ephi_elam_boundary}
\e^{-\lambda(R)}=\e^{\nu(R)}=1-\frac{2M}{R},
\end{equation}
suffice to determine the equilibrium configuration of a neutron
star of mass $M$ and the spacetime around it as well.

For a given physical EOS and a given central density
$\rho(r=0)=\rho_0$, this system of equations can be solved by
numerical integration. However, the problem can also tackled by
expanding relevant quantities as power series  of $r$ about the
origin \citep{Chandrasekhar1}:
\begin{eqnarray}
\rho(r)&=&\rho_0+\rho_2 r^2 +\ldots \,,\\
m(r)&=&  4\pi\left(\frac{\rho_0 r^3}{3} + \frac{\rho_2 r^5}{5} + \ldots \right)\,,\\
    P(r)&=&P_0+P_2 r^2+ \ldots \,.
\end{eqnarray}
Similarly, the series expansions of the metric coefficients are
readily obtainable from (\ref{g1}) and (\ref{g2}). Direct
substitution of these expansion into the TOV equations and the EOS
yields formulas expressing all the expansion coefficients in terms
of the central density $\rho_0$ \citep{Chandrasekhar1}.

Motivated by this expansion scheme that works nicely about the
origin $r=0$, we consider a mass distribution:
\begin{equation}
m_\c(r)= M \left[ \frac{5}{2}\left(\frac{r}{R}\right)^3
-\frac{3}{2}\left(\frac{r}{R}\right)^5 \right],
\end{equation}
which was first introduced by \citet{Tolman:1939jz} and is
commonly referred to as Tolman VII model (TVIIM). In addition to
being a good approximation near the origin, TVIIM satisfies the
boundary condition $\rho(r=R)=0$ at the stellar surface, which is
correct for most EOS as $P(r=R)=0$. Therefore, it is reasonable to
conjecture that TVIIM is able to provide a good global
approximation to the exact mass distribution. The conjecture is
indeed confirmed by the results shown in Fig.~\ref{mass_0.28_2},
where the normalized mass distribution function $m(r)/M$ is
plotted against the scaled radius $r/R$ for stars described by
different EOS with a common compactness $M/R=0.2$. It is manifest
that $m(r)$ obtained from stars with different EOS are close to
that of TVIIM, and similar conclusion has also been drawn recently
by \citet{Lattimer:2001}. This discovery suggests that the QNMs of
TVIIM could be close to those of realistic neutron stars.

Moreover, TVIIM has a distinct feature that gives rise to the
universal behaviours observed in the $w$-mode QNMs. Introducing
the scale transformations:
\begin{eqnarray}
\tr &=& r/M,  \\
\tm(\tr) &=& {m(r)}/{M},  \\
\tP(\tr) &=& M^2 P(r),  \\
\trho(\tr) &=& M^2 \rho(r),  \\
\tphi(\tr) &=& \nu(r),
\end{eqnarray}
we rewrite the TOV equations as follows:
\begin{eqnarray}\label{TOV}
\frac{\d\tm}{\d\tr} &=& 4\pi \tr^2\trho, \\
\frac{\d\tphi}{\d\tr} &=& \frac{2\tm+8\pi \tr^3 \tP}{\tr(\tr-2\tm)}, \\
\frac{\d\tP}{\d\tr} &=&
-\frac{1}{2}(\trho+\tP)\frac{\d\tphi}{\d\tr}.
\end{eqnarray}
These scaled TOV equations have exactly the same forms as the
original ones. In particular, for TVIIM the scaled mass
distribution function is given by:
\begin{equation}
\tm_\c(\tr)\equiv
m_\c(r)/M=\frac{5}{2}\left(\frac{M}{R}\right)^3\tr^3
-\frac{3}{2}\left(\frac{M}{R}\right)^5\tr^5.
\end{equation}
It is clear that the function $\tm_\c(\tr)$ is solely
characterized by the compactness $M/R$ and is independent of the
mass $M$. Consequently, other scaled functions (e.g. $\trho(\tr)$,
$\tP(\tr)$ and $\tphi(\tr)$) associated with TVIIM depend only on
$M/R$ as well. In fact, this statement holds as long as $\tm(\tr)$
does not contain any explicit mass-dependence. However, as shown
in Fig.~\ref{mass_0.28_2}, TVIIM provides the simplest, yet
robust, global approximation to $m(r)$. Besides, TOV equation for
TVIIM is exactly solvable \citep{Tolman:1939jz}, thus further
simplifying relevant calculations. In the subsequent discussion,
we will show that both the axial and the polar gravitational wave
equations also exhibit similar scaling behaviour, thus leading to
the universality in the QNMs.

\section{Scaling behaviour of axial QNMs}
The equation of motion for relativistic stars can be decomposed in
terms of spherical harmonics and in turn classified into axial and
polar oscillations \citep{Thorne67,Lindblom_1983,Chandrasekhar1}.
The equation for the axial oscillations of neutron stars is given
by a Regge-Wheeler-type equation \citep{Chandrasekhar1}:
\begin{equation}
\label{KG_eq} \left[\frac{\d^2}{\d
r_{*}^2}+\omega^2-V_{}(r_{*})\right]\psi_{}(r_{*}) = 0,
\end{equation}
where the tortoise coordinate $r_{*}$ and the potential $V$ are
defined by
\begin{equation}\label{r*_in} r_{*}=\int_{0}^r
\e^{(-\nu+\lambda)/2} dr,
\end{equation}
and
\begin{equation}\label{RW_V}
V_{}(r_{*})=\frac{\e^\nu}{r^3}[l(l+1)r+4\pi r^3(\rho-P)-6m(r)],
\end{equation}
respectively. Outside the star, the pressure and the density
vanish and hence the tortoise radial coordinate there can be
simplified to
\begin{equation}\label{r*_out}
r_{*}=r+2M \ln\Big(\frac{r}{2M}-1\Big)+C,
\end{equation}
where $C$ is a constant that can be obtained by matching
(\ref{r*_in}) with (\ref{r*_out}) at $r=R$. Similarly, the
potential $V$ is reduced to the well known Regge-Wheeler potential
\citep{RWeq}.
\begin{equation}\label{RWP_out}
V_{\rm
rw}(r_{*})=\left(1-\frac{2M}{r}\right)\left[\frac{l(l+1)}{r^2}-
\frac{6M}{r^3}\right].
\end{equation}

The  axial mode wave equation can be scaled by multiplying $M^2$
to (\ref{KG_eq}), yielding
\begin{equation}
\label{scale_KG_eq}
\left[\frac{\d^2}{\d\tx^2}+\tomega^2-\tV_{}(\tx)\right]\tpsi_{\a}(\tx)
= 0,
\end{equation}
where
\begin{eqnarray}
\tomega &=& M\omega,  \\
\tx &=& {r_*}/{M},  \\
\tpsi_\a(\tx) &=& \psi(r_*),
\end{eqnarray}
and
\begin{equation}
\tV_{}(\tx)=\left\{%
\begin{array}{ll}
    {\e^{\tphi}}\tr^{-3}\left[l(l+1)\tr+4\pi
\tr^3(\trho-\tP)-6\tm(\tr)\right], & \hbox{$\tr \leq R/M$;} \\
    \left(1-{2}{\tr}^{-1}\right)\left[{l(l+1)}{\tr^{-2}}-
{6}{\tr^{-3}}\right], & \hbox{$\tr > R/M$.} \\
\end{array}%
\right.
\end{equation}

\begin{figure}
\includegraphics[angle=270,width=8.5cm]{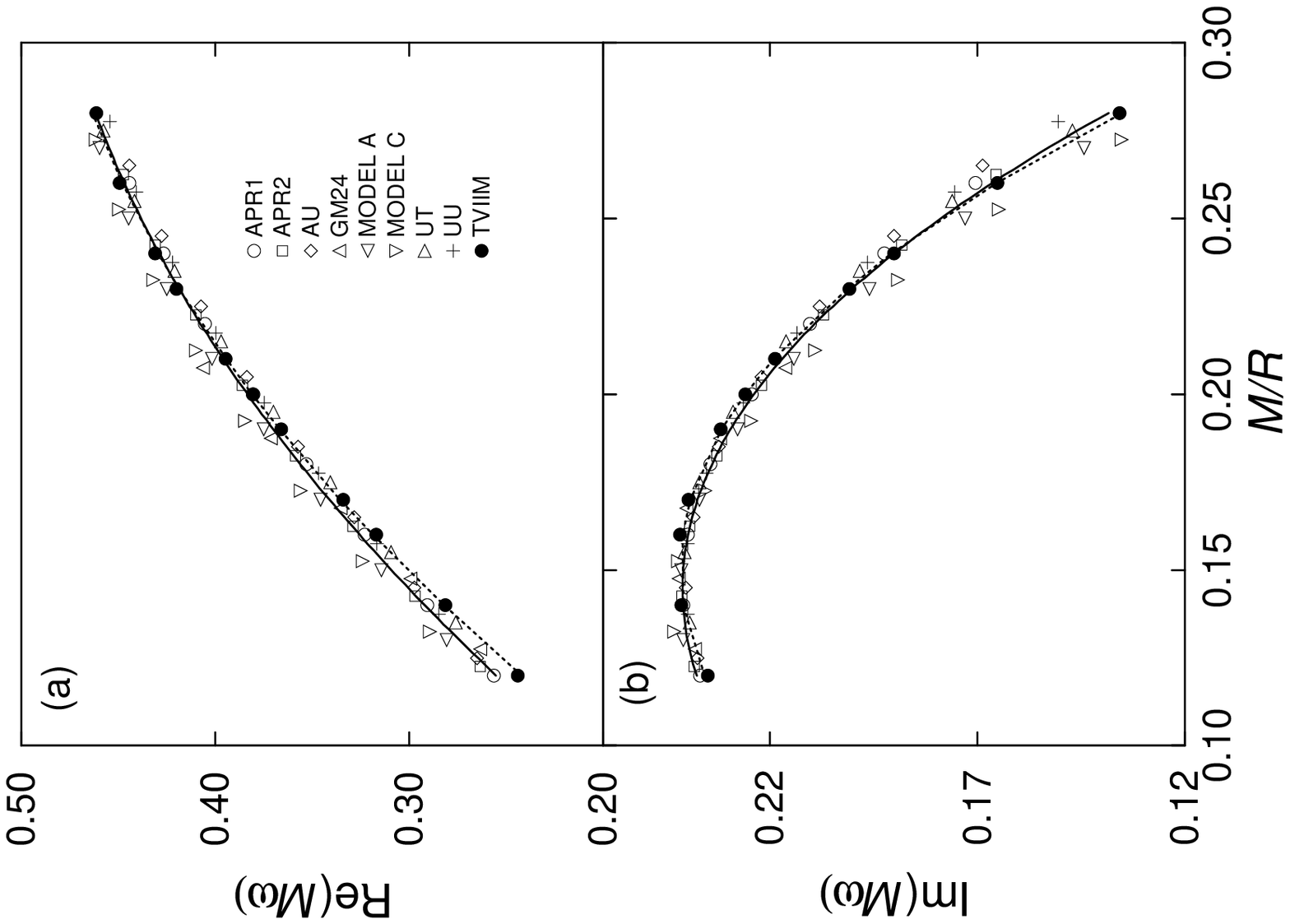}
\caption{(a) The real and (b) the imaginary parts of the scaled
frequency $M\omega$ of the least-damped axial $w$-mode are plotted
against the compactness $M/R$ for neutron stars described by
different EOS, including APR1, APR2, AU, GM24, Models A and C, UT,
UU, and also TVIIM. The solid line represents the best quadratic
fit in $M/R$ to the scaled frequencies of the eight realistic
stars. Likewise, the dotted-line is the best quadratic fit to
those of TVIIM.} \label{ldw_a}
\end{figure}

\begin{figure}
\includegraphics[angle=270,width=8.5cm]{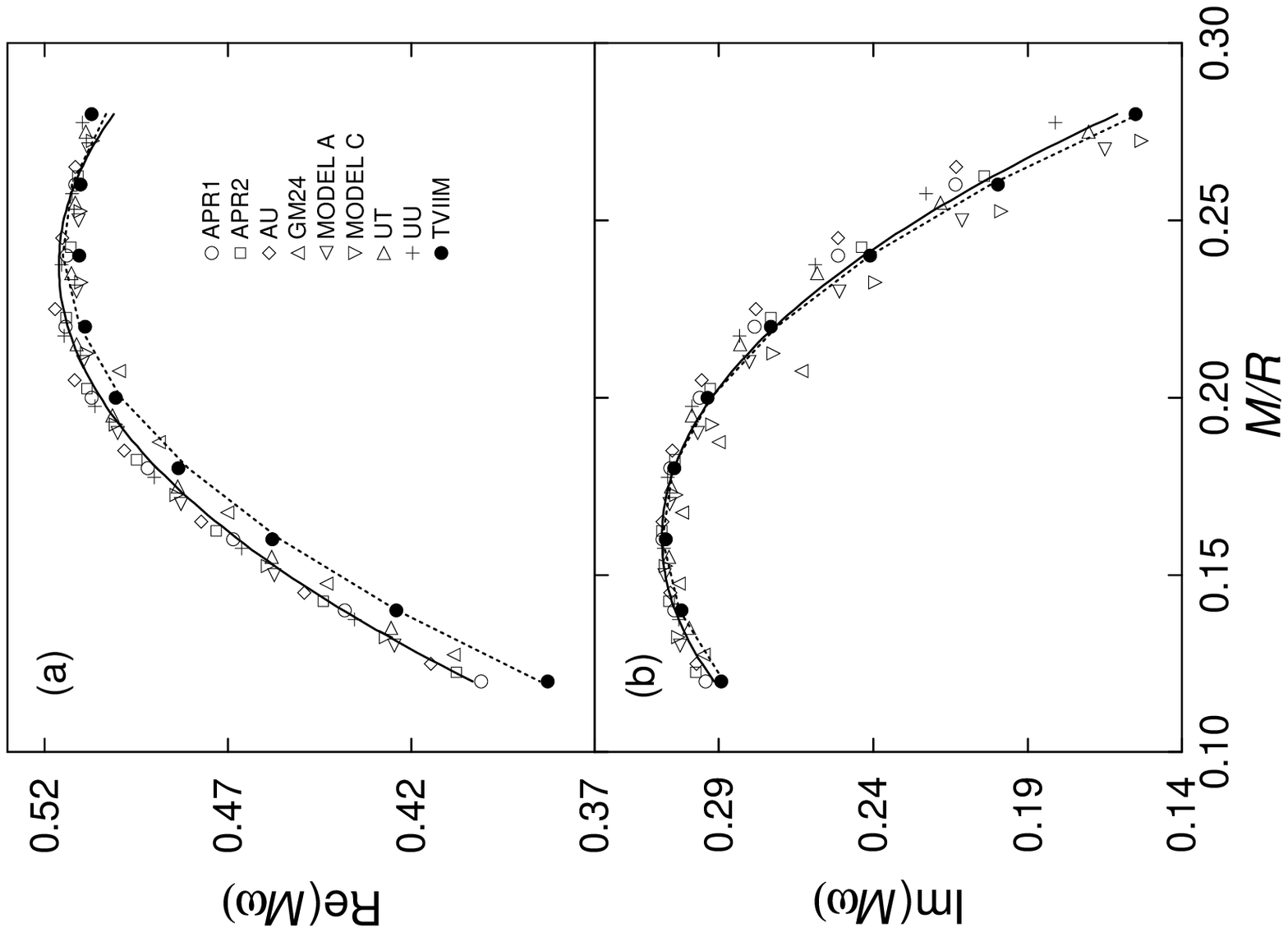}
\caption{(a) The real and (b) the imaginary parts of the scaled
frequency $M\omega$ of the least-damped polar $w$-mode are plotted
against the compactness $M/R$ for neutron stars described by
different EOS (see the caption of Fig.~1). The solid and
dotted-lines represent the best quadratic fit in $M/R$ to the
scaled frequencies of the eight realistic stars and TVIIM,
respectively.} \label{ldw_p}
\end{figure}

\begin{figure}
\includegraphics[angle=270,width=8.5cm]{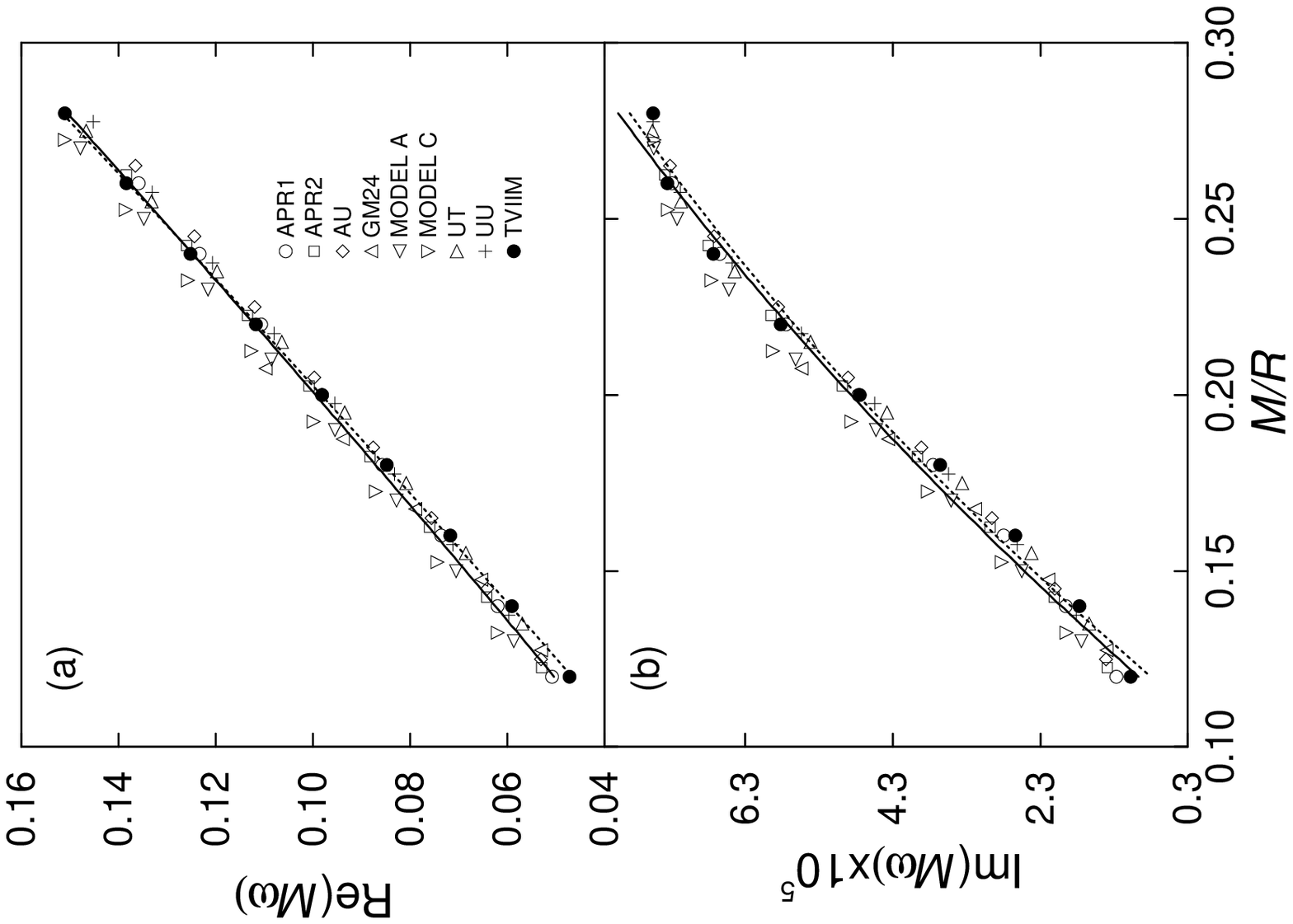}
\caption{(a) The real and (b) the imaginary parts of the scaled
frequency $M\omega$ of the fluid $f$-mode are plotted against the
compactness $M/R$ for neutron stars described by different EOS
(see the caption of Fig.~1). The solid and dotted-lines represent
the best quadratic fit in $M/R$ to the scaled frequencies of the
eight realistic stars and TVIIM, respectively.} \label{f_mode}
\end{figure}

\begin{figure}
\includegraphics[angle=270,width=8.5cm]{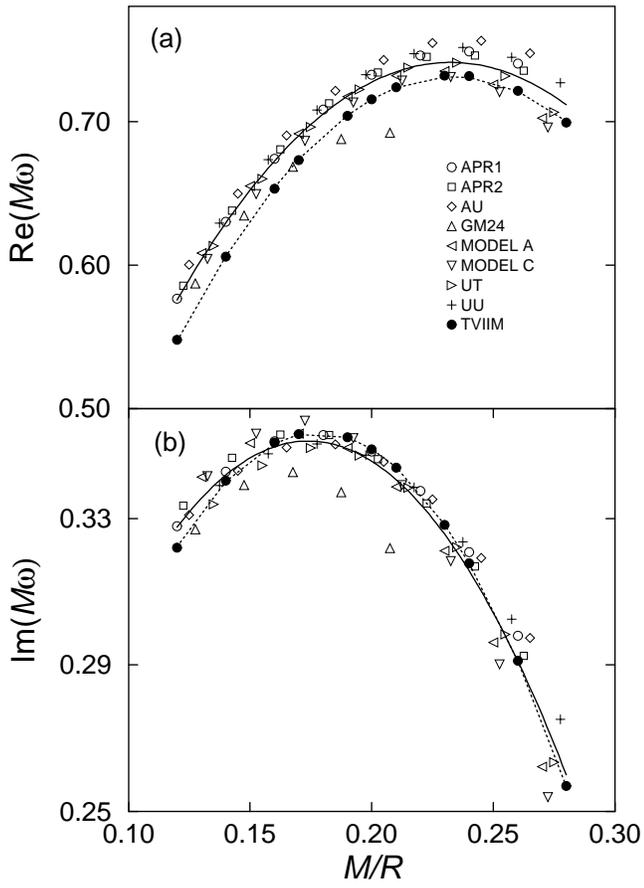}
\caption{Same as Fig.~1 for the second-least damped axial
$w$-mode.} \label{ldw_a2}
\end{figure}

\begin{figure}
\includegraphics[angle=270,width=8.5cm]{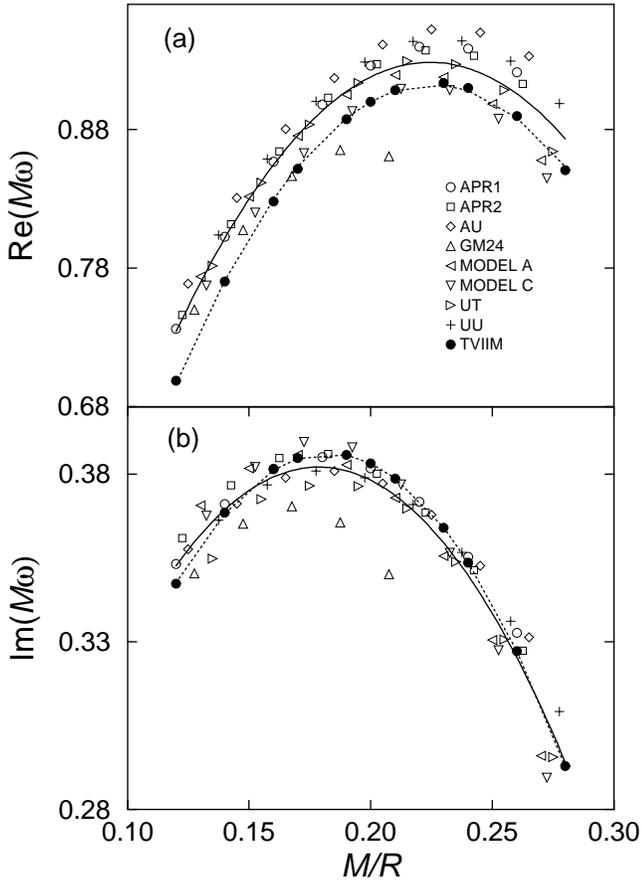}
\caption{Same as Fig.~2 for the second-least damped polar
$w$-mode.} \label{ldw_p2}
\end{figure}

\begin{figure}
\includegraphics[angle=270,width=8.5cm]{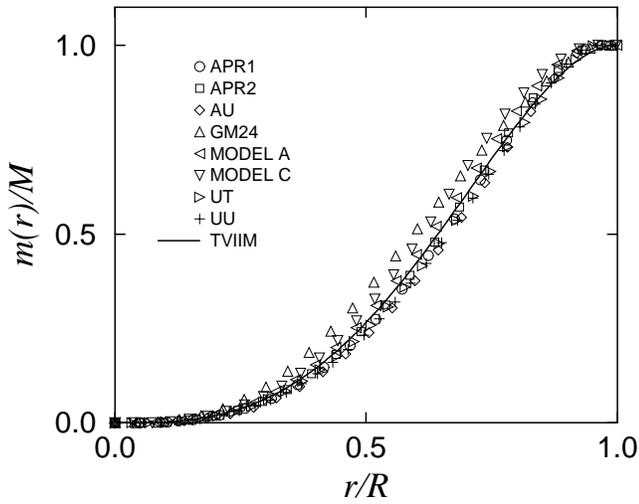}
\caption{The mass distribution functions of  neutron stars
characterized by eight realistic different EOS and  with a common
compactness $M/R=0.2$ are shown. The mass distribution function of
TVIIM, the solid line, is shown as a benchmark for realistic
stars.} \label{mass_0.28_2}
\end{figure}

For TVIIM, $\tm(\tr)=\tm_\c(\tr)$, $\trho$ and $\tP$ are universal
functions of $\tr$ with the compactness being the unique
parameter. As a consequence, QNMs of TVIIM  depend only on the
compactness. For other realistic stars whose mass profiles are
similar to that of TVIIM, it is plausible to expect that the
scaled QNM frequency $\tomega$ is close to that of TVIIM with the
same compactness. Hence, we argue that $\tomega$ of realistic
stars has a trend similar to that of TVIIM. This conjecture is
confirmed in Fig.~\ref{ldw_a}, where  QNMs of TVIIM
(solid-circles) are shown to be close to those of other realistic
stars. It is amazing that the dotted line, the best quadratic fit
to QNMs of TVIIM, almost coincides with the solid line that is
obtained by fitting a quadratic expression in $M/R$ to $\tomega$
for stars with eight different EOS. Obviously, the universal
behaviour in QNMs of neutron stars, including realistic ones and
TVIIM, is well approximated by the dotted line. For reference and
comparison, we record the values of the regression parameter $a$,
$b$ and $c$ (see (\ref{regression})) of the dotted line in Table
1. One can see that the parameters $a$ and $b$ of the two lines
are indeed quite close, which agrees with our observation from
Fig.~\ref{ldw_a}. On the other hand, there is a disparity between
the parameters $c$ for the two cases. However, such a disparity is
obviously unimportant and does not affect the agreement between
the two curves.
\section{Scaling behaviour of polar QNMs}
The equation of motion for polar case allows pulsating modes of
various kinds, including the fundamental fluid $f$-mode, the
gravitational wave $w$-mode, the pressure $p$-mode and the gravity
$g$-mode \citep[see, e.g, a recent review by][]{Kokkotas_rev}. To
investigate why QNMs of $f$-mode and polar $w$-mode oscillations
display generic behaviour as discussed above, we adopt the
notation of \citet{Lindblom_1983} and consider the set of polar
wave equations governing the variation of the three metric
perturbation functions, $H_{0}$, $H_{1}$, and $K$, and the two
fluid perturbation functions, $W$ and $V$, inside the star ($r \le
R$) \citep{Lindblom_1983}:
\begin{eqnarray}
 H_{1}^{\prime} & = & - {1 \over r} \left[ l + 1 + {{2 m \e^{\lambda}}\over {r}}
 + 4 \pi r^{2} \e^{\lambda} (P - \rho) \right] H_{1} + {1 \over r} {\e^{\lambda}}
{[H_{0} + K - 16 \pi (\rho + P) V]}, \label{polar_DE1}\\
 K ^{\prime} & = & {1 \over r}H_{0} + {1 \over {2 r}} l (l + 1) H_{1}-
\left[  {{(l + 1)} \over r} - { {\nu^{\prime}} \over 2} \right] K
-
{{8 \pi (\rho + P) \e^{\lambda/2} }\over r} W, \label{polar_DE2}\\
 W ^{\prime} & = & - {{(l + 1)} \over r} W + r \e^{\lambda/2} \left[
 {{\e^{-\nu/2}}\over {\gamma P}} X - {{l (l + 1)}\over {r^{2}}} V
 + {1 \over 2} H_{0} + K \right] , \label{polar_DE3}\\
X ^{\prime} & = & - {l \over r} X + (\rho + P) \e^{\nu/2} \left \{
{1 \over 2} {\left({1 \over r} - {{\nu^{\prime}}\over 2} \right)}
H_{0} + {1 \over 2} {\left[ r \omega^{2} \e^{-\nu} +
 {1 \over 2} {{l (l + 1)} \over {r}} \right]} H_{1} \right. +
 {1 \over 2} \left({{3 \over 2} {\nu^{\prime}} - {1 \over r}}
\right) K - {1 \over 2}{{ l (l + 1)}\over {r ^{2}}} {\nu^{\prime}}
V\nonumber -\nonumber \\ & & \left. {1 \over r} \left[ 4 \pi (\rho
+ P) \e^{\lambda/2} + \omega^{2} \e^{\lambda/2 - \nu} - {1 \over
2} r^{2}
 {\left({ {\e^{-\lambda/2}{\nu^{\prime}} }\over {r^{2}}}\right)}^{\prime}
\right] W \right\} . \label{polar_DE4}
\end{eqnarray}
Here $\gamma$ is the adiabatic index:
\begin{equation}
\gamma = \left(1+\frac{\rho}{P}\right)\frac{\d P}{\d\rho},
\end{equation}
and $X$ is defined as:
\begin{eqnarray}\label{defintion_X}
X &= & \omega^{2} (\rho + P) \e^{-\nu/2} V - {1 \over r}
P^{\prime} \e^{(\nu-\lambda)/2} W + {1 \over 2} (\rho + P)
\e^{\nu/2} H_{0},
\end{eqnarray}
which is related to the Lagrangian perturbation in pressure,
$\Delta p$, through the equation $X=-\e^{\nu/2}\Delta p$
\citep{Lindblom_1983}. These four first order ODE's, together with
the constraint equation, completely determine the dynamics inside
the star.

Outside the star, the whole problem could be described by only two
metric perturbation functions, namely $H_{0}$ and $K$. A linear
combination of these two functions defines the Zerilli wave
function $Z(r_{*})$ that satisfies the Zerilli wave equation
\citep{Zerilli}:
\begin{equation}
\label{KG_eq_Z} \left[\frac{\d^2}{\d r_{*}^2}+\omega^2- V_{\rm
Z}(r_{*})\right] Z(r_{*}) = 0,
\end{equation}
with $V_{\rm Z}(r_{*})$ being the Zerilli potential,
\begin{equation}
 V_{\rm Z}(r_{*}) = {{1-2 M/r}\over{r^{3} {(n r + 3 M )}^{2}}}
\left[ 2n^{2}(n+1)r^{3}+ 6 n^{2}M r^{2} + 18 n M^{2} r + 18 M^{3}
\right],
\end{equation}
and $n = (l-1)(l+2)/2$.

\begin{figure}
\includegraphics[angle=270,width=8.5cm]{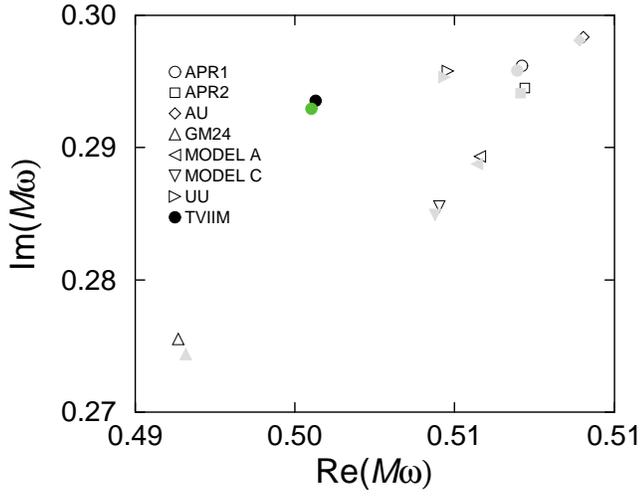}
 \caption{The complex eigenfrequencies of the least damped
 polar $w$-mode for stars constructed from seven realistic
 EOS (unfilled symbols) and TVIIM (solid-circle), which have a common compactness $M/R=0.2$, are shown.
 The grey (or dark-grey for the case of TVIIM) symbols indicate the
 corresponding results obtained under the approximation $X=0$.}
\label{zeroX}
\end{figure}

A close examination of the polar wave equations inside the star
reveals that the scaling behaviour is likely to be marred. The
culprit is the presence of the adiabatic index $\gamma$, which
directly depends on $P$ (instead of $\tilde{P}$) and is hence not
amenable to scaling. In fact, it has been shown QNMs of $p$-mode
oscillations (especially the damping rate) do not follow the
universal behaviour. However, as $X$ is proportional to the
Lagrangian perturbation in the pressure, it should be negligibly
small for $w$-mode and $f$-mode waves. To verify this issue, we
solved the polar wave equations with the assumption $X=0$ and
found that eigenfrequencies of $w$-mode and $f$-mode waves are
almost unaffected by such assumption (see Fig.~\ref{zeroX} for the
result in the $w$-mode case). This shows that the term $X/\gamma$
in (\ref{polar_DE3}) can be omitted as far as $w$-mode and
$f$-mode waves are concerned.

Under the approximation $X=0$, which is valid for $w$-mode and
$f$-mode oscillations, we can show that the universality observed
in the axial case also exists in the polar case. Again we have to
introduce more scale transformations:
\begin{eqnarray}
 \tW&=&W/M^2, \\
\tX&=&M^2X, \\
\tV&=&V/M^2,
\end{eqnarray}
while for $H_0$, $H_1$, $K$, $\lambda$,  the transformed
quantities are identical to the original ones. In terms of these
scaled variables, Eqs.~(\ref{polar_DE1}), (\ref{polar_DE2}) and
(\ref{polar_DE3}) then become
\begin{eqnarray}
\tHI^{\prime} & = & - {1 \over \tr} \left[ l + 1 + {{2 \tm
\e^{\tlam}}\over {\tr}}
 + 4 \pi \tr^{2} \e^{\tlam} (\tP - \trho) \right] \tHI + {1 \over \tr} {\e^{\tlam}}
{[\tHO + \tK - 16 \pi (\trho + \tP) \tV]}, \\
 \tK ^{\prime} & = & {1 \over \tr}\tHO + {1 \over {2 \tr}} l (l + 1) \tHI-
\left[  {{(l + 1)} \over \tr} - { {\tphi^{\prime}} \over 2}
\right] \tK -
{{8 \pi (\trho + \tP) \e^{\tlam/2} }\over \tr} \tW, \\
 \tW ^{\prime} & = & - {{(l + 1)} \over \tr} \tW + \tr \e^{\tlam/2} \left[
  - {{l (l + 1)}\over {\tr^{2}}} \tV
 + {1 \over 2} \tHO + \tK \right] ,
\end{eqnarray}
where a prime in these transformed equations  indicates
differentiation with respect to $\tr$. Note that there is no need
to consider (\ref{polar_DE4}) because of the approximation $X=0$.
Besides, Eq.~(\ref{defintion_X})  can be rewritten accordingly and
manifest scale invariance.

Similarly, we define the transformed Zerilli wave function
$\tilde{Z}=Z /M$, which satisfies a scaled wave equation outside
the star:
\begin{equation}
\label{scale_KG_eq_Z} \left[\frac{\d^2}{\d\tx^2}+\tomega^2-
\tV_{\rm Z}(\tx)\right]\tilde{Z}(\tx) = 0,
\end{equation}
with the scaled potential $\tV_{\rm Z}(\tx)$ given by:
\begin{equation}
\tV_{\rm Z}(\tx) = {{1-2/\tr}\over{\tr^{3} {(n \tr + 3 )}^{2}}}
\left[ 2n^{2}(n+1)\tr^{3}+ 6 n^{2} \tr^{2} + 18 n \tr + 18
\right].
\end{equation}

It is clear that the scaled polar oscillation equations are
explicitly mass-independent. Hence, the scaled polar QNM
frequencies $\tomega$ of TVIIM could also display a similar
universality as those of the axial oscillations, save cases in
which perturbation in pressure becomes non-negligible, e.g. the
pressure $p$-mode. Since the mass profiles of other realistic
stars are close to that of TVIIM, their QNMs should follow the
same trend. In Fig.~\ref{ldw_p} and Fig.~\ref{f_mode}, the solid
circles show the scaled frequencies of polar $w$-mode and $f$-mode
QNMs of TVIIM, which clearly illustrate the scaling behaviour of
these modes. Besides, they are also well approximated by the solid
(dotted) line obtained from the best quadratic fit to QNMs of
realistic neutron stars (TVIIM). The similarity between these
lines in turn establishes independent corroboration of our theory.

We have shown here that the scaled polar QNM frequencies $\tomega$
of realistic neutron stars are approximately given by a universal
function of the compactness. Our finding is particularly
interesting for the $f$-mode oscillation because Eq.~(\ref{TL}) is
different from the empirical formulas (\ref{BBF_f1}) and
(\ref{BBF_f2}) discovered previously \citep{Andersson1998}.  As
mentioned in Sect.~I, Eq.~(\ref{BBF_f1}) in general leads to
explicit mass-dependence of $M\omega_\r$ and is inconsistent with
the scaling behaviour of the polar oscillation equation. Moreover,
comparing Fig.~3 in the present paper with Fig.~1 in the paper by
\citet{Andersson1998}, we conclude that Eq.~(\ref{TL}) can indeed
provide a better fit to the scaled oscillation frequency
$M\omega_\r$. On the other hand, despite that Eq.~(\ref{BBF_f2})
is apparently different from (\ref{TL}), it implies that
$M\omega_\i=M/\tau$ is still a function of the compactness,
namely:
\begin{eqnarray}
M\omega_\i  \approx
\left(\frac{M}{R}\right)^4\left[\alpha_\i\left(\frac{M}{R}\right)+\beta_\i
\right].\label{BBF_f2_M}
\end{eqnarray}
Hence, it is again in perfect agreement with the scaling behaviour
of the polar oscillation equation. It is worthy of remark that the
leading dependence in (\ref{BBF_f2_M}), $M\omega_\i \propto
({M}/{R})^4$, can be argued from the quadrupole radiation formula
\citep{Andersson1998} and is expected to be a good approximation
as long as the compactnesss is small. In other words,
Eq.~(\ref{BBF_f2_M}) is a Taylor expansion of $M\omega_\i$ about
the point ${M}/{R}=0$, whereas  Eq.~(\ref{TL}) expands
$M\omega_\i$ about a non-zero value of ${M}/{R}$ (say,
${M}/{R}\approx 0.2$ in the current situation). Therefore, the
disparity between (\ref{TL}) and (\ref{BBF_f2_M}) is reconciled.
As the compactness of realistic neutron stars usually lies in the
range of $0.1 - 0.3$, we expect that these two formulas can
complement each other in the prediction of QNM frequencies of
$f$-mode oscillations.
\section{Application}
\begin{figure}
\hspace{0.9cm}
\includegraphics[angle=270,width=7.5cm]{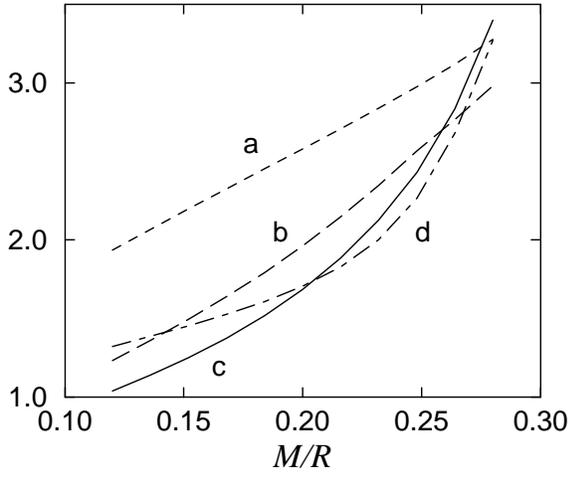}
 \caption{(a) $10\times{\rm Re}
\omega_{\f}/{\rm Re} \omega_{\a}$ (short dashed-line);  (b)
$10\times{\rm Re} \omega_{\f}/{\rm Re} \omega_{\p}$ (long
dashed-line); (c) ${\rm Re} \omega_{\a}/{\rm Im} \omega_{\a}$
(solid-line); and (d) ${\rm Re} \omega_{\p}/{\rm Im} \omega_{\p} $
(dot-dashed-line) are plotted against the compactness $M/R$.}
\label{mono}
\end{figure}

As TVIIM successfully captures the universality in QNMs of $w$-
and $f$-modes, it can be used to set up a benchmark for realistic
stars. Bases on this, a viable scheme to infer the mass and the
radius of a neutron star from its $w$- and $f$-modes QNMs is
proposed here. In Fig.~\ref{mono}, we plot $10\times{\rm Re}
\omega_{\f}/{\rm Re} \omega_{\a}$ (short dashed-line);
$10\times{\rm Re} \omega_{\f}/{\rm Re} \omega_{\p}$ (long
dashed-line); ${\rm Re} \omega_{\a}/{\rm Im} \omega_{\a}$
(solid-line); and ${\rm Re} \omega_{\p}/{\rm Im} \omega_{\p} $
(dot-dashed-line)  against the compactness, where $\omega_\a$,
$\omega_\p$ and $\omega_\f$ are QNM frequencies of the leading
axial and polar $w$-modes, and $f$-mode of TVIIM, respectively.
The curves are all monotonically increasing function of $M/R$ in
the regime $0.1 <M/R <0.3$ where physical neutron stars are
stable. Depending on experimental data available, we can use any
one of these ratios to approximately infer the  compactness of an
unknown neutron star. Once the compactness is known, we can
readily read the approximate mass of the star from Figs.~$1-5$ if
the real (or imaginary) part of relevant QNM frequency is obtained
in gravitational wave detection. Since both the real and imaginary
parts of the scaled frequencies of $w$- and $f$-modes are well
approximated by quadratic functions of $M/R$, implementation of
the above-mentioned procedure is straightforward. Still,
feasibility of this scheme hinges on availability of accurate
experimental data, which is beyond the scope of the present paper
\citep[see, e.g.][for a discussion on this issue]{Kokkotas_2001} .

Here we illustrate our method with the following example. Consider
a neutron star constructed from APR2 model \citep{APR}, whose mass
and radius are $1.596 {\rm M}_\odot$ and 11.77~km, respectively.
The first polar $w$-mode QNM of this star has a angular frequency
64.57~kHz and a damping time of $2.666 \times 10^{-5}$~s.
Therefore, the ratio ${\rm Re} \omega_{\p}/{\rm Im} \omega_{\p} $
is equal to 1.722. If such wave signals are detected by us, we can
use TVIIM as a benchmark for this star and infer its approximate
mass and radius from the wave data. From the ratio ${\rm Re}
\omega_{\p}/{\rm Im} \omega_{\p}=1.722 $ and the dot-dashed-line
in Fig.~\ref{mono}, the compactness of the corresponding TVIIM
star is found to be 0.2025 (in geometrized units). Hence, the
scaled frequency can be obtained from Fig.~\ref{ldw_p}. By
comparing the scaled frequency with unscaled one, the mass of the
star can be readily found, which is $1.581 {\rm M}_\odot$ in this
case. Similarly, the radius of the star can be found from the
compactness given above and is equal to 11.51 km. Both the values
of the radius and the mass inferred from TVIIM star are close to
those of APR2 star. From this example one can see that TVIIM can
in fact serve as a template for other realistic stars. In this
regard, the role of TVIIM in gravitational wave analysis is
reminiscent of that of the hydrogen atom in atomic spectroscopy.

\begin{figure}
\includegraphics[angle=270,width=8.5cm]{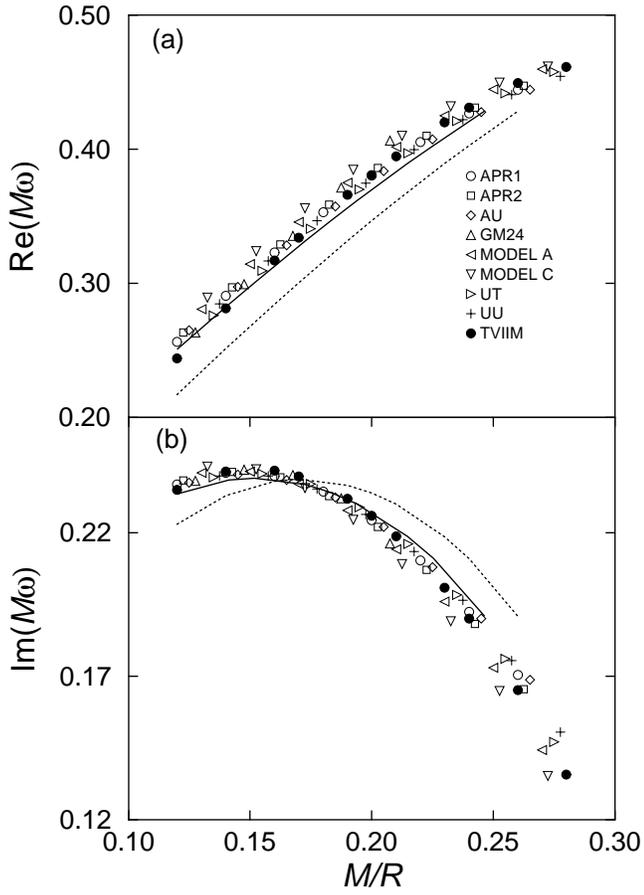}
 \caption{(a) The real and (b) the imaginary parts of the scaled
frequency $M\omega$ of the least-damped polar $w$-mode are plotted
against the compactness $M/R$ for neutron stars described by
different EOS (see caption of Fig.~1). The solid and dotted-lines
respectively show the results of ``dressed" and bare strange quark
stars.} \label{qstar}
\end{figure}

 Before we can put this scheme into practice, we have
to further examine whether the universality demonstrated in
Figs.~$1-5$ is truly generic. As discussed above, the universality
is closely tied with the validity of TVIIM, which is developed
under the assumption that the density function $\rho(r)$ is
expandable in series of $r$. Despite that the assumption is
appealing, there is one well-known exception to it, namely the
strange star \citep{Witten_84}. One salient feature of the EOS of
strange quark matter, say the MIT bag model \citep[see,
e.g.][chap.~8]{ComStar}, is that the density is non-zero at zero
pressure. Therefore, the density of a bare quark star (i.e. one
without a normal nuclear matter shell) does not vanish at the
surface of the star. Obviously, TVIIM fails to mimic such stars as
there is a discontinuity in the density on the stellar surface. As
shown in Fig.~\ref{qstar}, the QNMs of bare strange quark stars
deviate marked from the university demonstrated by other stars
that are described by continuous density profiles. To pinpoint the
cause of the deviation,  the QNMs of strange quark stars with a
normal nuclear matter shell are also shown in Fig.~\ref{qstar}. It
is obvious the QNMs of such hypothetically ``dressed" quark star,
which have no discontinuities in their density profiles, also
display the universality discussed in the present paper. We
conclude that the scheme proposed above works properly as long as
the star in consideration is characterized by a continuous density
profile. Furthermore, by virtue of the TOV equations, the pressure
is always a continuous function of $r$ and the density profile is
continuous if relevant EOS $P(\rho)$ is a continuous function of
$\rho$. As a result, deviation from the universal curve might
indicate a discontinuous (or rapidly varying) EOS. This effect can
indeed be observed from the QNM frequencies of GM24 shown in
Fig.~\ref{ldw_a2} and \ref{ldw_p2}. Neutron stars constructed from
GM24 will become unstable if $M/R >0.21$ owing to emergence of
phase transition that leads to discontinuous (or rapidly varying)
EOS. Correspondingly, QNM frequencies of GM24 stars no longer
follow the those of TVIIM as its compactness approaches that
limit.
\section{Conclusion}
In this paper we have unveiled some universal scaling behaviours
prevailing in QNM frequencies of $w$- and $f$-modes. In addition,
we have shown that these behaviours are attributable to the
generic form of the mass distribution function. Based on our
discovery, a scheme is proposed here to infer the mass and the
radius of an unknown star from its gravitational wave spectrum.
Despite that there might be some exceptional cases, e.g. bare
strange stars, the scheme does provide reasonable estimate of
relevant values in normal situation. As it has been shown that the
mass and the radius of a neutron star can be used to infer the EOS
of stellar matter \citep{Lindblom_invert}, our discussion will
also lead to determination of EOS from gravitational wave signals.

The TVIIM considered here is a two-parameter model that is
characterized by the mass and the radius of a star. While
reproducing the main features of realistic stars, it cannot
capture the fine details in individual cases. Should it be
possible to gather more than one QNM frequency of a pulsating
star, we could, in principle, construct more refined template for
it. We are currently working along this direction and relevant
result will be reported in due course.

 Our analysis sketched above is based on assumption that QNMs
of gravitational waves are clearly discernible. However, it is
well known that gravitational wave signals are usually overwhelmed
by noises of various kinds, including thermal, quantum  and
gravity-gradient noises \citep[see, e.g.][and references
therein]{Liu:2000ac,Thorne:1998hq,Santamore:2001nr}. As yet there
are no viable methods to completely eliminate these noises such
that clear and reliable gravitational wave signals can be
recorded. However, with the continuous efforts of researchers in
this field, we will surely get rid of these noisy gremlins at a
later date. Thus, gravitational wave astronomy will open up a new
channel to observe distant compact stars and the scheme outlined
here will be applicable in the near future.
\section*{Acknowledgments}
We thank J~Wu for discussions and assistance in preparation of the
manuscript. Our work is supported in part by the Hong Kong
Research Grants Council (grant No: CUHK4282/00P) and a direct
grant (Project ID: 2060260) from the Chinese University of Hong
Kong.
\newcommand{\noopsort}[1]{} \newcommand{\printfirst}[2]{#1}
  \newcommand{\singleletter}[1]{#1} \newcommand{\switchargs}[2]{#2#1}

\newpage

\newpage 


\newpage
\end{document}